\documentclass[aps,prl,twocolumn,showpacs]{revtex4}
\usepackage{amsmath,epsfig}
\begin{document}

\def \bv{\noindent{\bf V}}
\def \bg{\noindent{\bf g}}
\def \bs{\noindent{\bf s}}
\def \bx{\noindent{\bf \xi}} 
\def\bb{\noindent{\bf b}}
\def\bom{\noindent{\bf\Omega}}

\def \cv{{\cal V}}
\def \cg{{\cal G}}
\def \cs{{\cal S}}
\def \cx{{\cal X}} 
\def\cb{{\cal B}}
\def \cd{{\cal D}}
\def \com{{\cal \Omega}}

\def \sv{{\small V}}
\def \sw{{\small W}}

\title{Gradient learning in spiking neural networks by dynamic perturbation of conductances}

\author{Ila R. Fiete$^{1}$ and H. Sebastian Seung$^{2}$}
\affiliation{$^1$Kavli Institute for Theoretical Physics, University of California, Santa Barbara, CA 93106\\
$^2$Howard Hughes Medical Institute and Department of Brain and Cognitive Sciences, \\ M.I.T., Cambridge, MA 02139}

\begin{abstract} 
We present a method of estimating the gradient of an objective function with respect to the synaptic weights of a spiking neural network. The method works by measuring the fluctuations in the objective function in response to dynamic perturbation of the membrane conductances of the neurons. It is compatible with recurrent networks of conductance-based model neurons with dynamic synapses. The method can be interpreted as a biologically plausible synaptic learning rule, if the dynamic perturbations are generated by a special class of ``empiric'' synapses driven by random spike trains from an external source.
\end{abstract}

\maketitle



Neural network learning is often formulated in terms of an objective function that quantifies performance at a desired computational task. The network is trained by estimating the gradient of the objective function with respect to synaptic weights, and then changing the weights in the direction of the gradient. 

If neural and network dynamics and the objective function are all exactly known functions of the weights, such learning can be accomplished by explicitly computing the relevant gradients. A famous example of this approach, used with wide success in non-spiking, deterministic artificial neural networks \cite{LeCun98a}, is the {\em backpropagation} (BP) \cite{Rumelhart86, Widrow90} algorithm.  

However, the relevance of BP to neurobiological learning is limited. Biological neural activity can be noisy, and involves the highly nonlinear and often history-dependent dynamics of membrane voltages and conductances: neurons generate voltage spikes, and the efficacy of synaptic transmission varies dynamically, on a spike by spike basis \cite{Thomson94, Markram96}. Further, the objective function in neurobiological learning may depend on the dynamics of muscles and external variables of the world unknown to the brain. 
Similar complications are also present in analog on-chip or robotic implementations of machine learning.

For learning in such systems, alternative strategies are necessary. The method of {\em weight perturbation} estimates the gradients by perturbing synaptic weights, and observing the change in the objective function. Unlike BP, weight perturbation is completely ``model-free'' \cite{Dembo90} -- it does not depend on knowing anything about the functional dependence of the objective on the network weights -- and can be applied to stochastic spiking networks \cite{Seung03}. The disadvantage of a completely model-free approach is the tradeoff between generality and learning speed: weight perturbation is far more widely applicable than BP, but BP is much faster when it is applicable.

Here we propose a method that is intermediate between these two extremes, yet is applicable to arbitrary spiking neural networks. Instead of making perturbations to the synaptic weights, it estimates the $N^2$ weight gradients through dynamic perturbation of the conductances of the $N$ network neurons. Our algorithm does this by exploiting a feature generic to many models of neural networks: that inputs to a neuron combine additively before being subjected to further nonlinearities. Otherwise, the algorithm is model-free. Our approach generalizes the concept of {\em node perturbation}, which has been proposed for training feedforward networks of nonspiking neurons \cite{LeCun89,Widrow90} and can be much faster than weight perturbation \cite{Werfel05}.  We show how neural conductance perturbations can be biologically plausibly used to perform synaptic gradient learning in fully recurrent networks of realistic spiking neurons.


\noindent{\bf Spiking neural networks}\ We briefly discuss the mathematical conditions under which our assumption, that the synaptic inputs to a single neuron combine linearly, holds in spiking neural networks. If each neuron $i$ is electrotonically compact, it can be described by a transmembrane voltage $V_i$, obeying the current balance equation $C_i dV_i/dt = - I^{int}_i(t) - I^{syn}_i(t)$. The intrinsic current $I^{int}_i$ is generally a nonlinear function of voltage and dynamical variables associated with the spike-generating conductances in the membrane. The dynamics of these variables may be arbitrarily complex (e.g. Hodgkin-Huxley model) without affecting our derivations. A simple model for the synaptic current is $I^{syn}_i = \sum_j W_{ij} s_{ij}(t)(V_i(t)-E_{ij})$. The time-varying synaptic conductance from neuron $j$ to neuron $i$ is $W_{ij}s_{ij}(t)$, with amplitude controlled by the parameter $W_{ij}$. Its time course is determined by $s_{ij}(t)$, which could include complex forms of short-term depression and facilitation.  
If the reversal potentials $E_{ij}$ of the synapses are all the same, then the synaptic current can be written as $I^{syn}_i = g_i(t)(V_i(t)-E^{syn})$, where
\begin{equation}\label{eq:Linear}
g_i(t) = \sum_j W_{ij}s_{ij}(t)
\end{equation}
is the sum of all postsynaptic conductances of the synapses onto neuron $i$. The linear dependence of $g_i(t)$ on the synaptic weights $W_{ij}$ will be critical below. However, this linear dependence may be  embedded inside a nonlinear network, which may be arbitrarily complex without afffecting the following derivations. In fact, all networks -- neural and spiking or neither -- that depend on a set of interaction variables $s_{ij}(t)$ and parameters $W_{ij}$ through Eq. (\ref{eq:Linear}) satisfy the necessary conditions for our derivation below. 

\noindent{\bf Gradient learning}\  
We represent the state of the network by a vector $\Omega(t)$, which includes the synaptic variables $s_{ij}(t)$ and all other dynamical variables (e.g.,  the voltages $V_i(t)$ and all variables ssociated with the membrane conductances). Starting from an initial condition $\Omega(0)$ the network generates a trajectory from time $t=0$ to $t=T$, and in response receives a scalar ``reinforcement'' signal $R[\Omega]$, which is an arbitrary functional of the trajectory. For now we assume that the network dynamics are deterministic, and present the fully stochastic case in the Appendix. Each trajectory along with its reinforcement is called a ``trial,'' and the learning process is iterative, extending over a series of trials. The signal $R$ depends implicitly on the synaptic weights $W_{ij}$, and is an objective function for learning. In other words, the goal of learning is to find synaptic weights that maximize $R$. A heuristic method for doing this is to follow the gradient of $R$ with respect to $W_{ij}$. Next we derive our gradient learning rule. 
\begin{figure}[h]
\centerline{\epsfig{figure=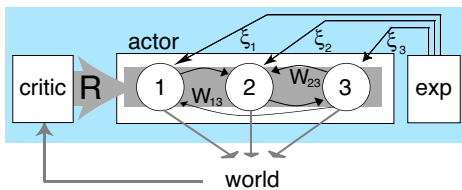,height = 1in}}
\caption[The model network.]{Neurons in a recurrent network (``actor''), connected by modifiable weights $W$. In addition, each neuron $i$ receives an empiric synapse carrying perturbing input $\xi_i(t)$ from an external ``experimenter''. A global reinforcement signal R is broadcast by a ``critic'' to all neurons in the network.}
\label{fig:network}
\end{figure}

\noindent{\bf Sensitivity lemma}\
Suppose that $W_{ij}(t)$ were a time-varying function.
Then by Eq.\ (\ref{eq:Linear}) and the chain rule, it would follow that
\begin{equation}
\frac{\delta R}{\delta W_{ij}(t)} = \frac{\delta R}{\delta g_i(t)}s_{ij}(t)
\end{equation}
But if $W_{ij}(t)$ is constrained to take on the same value at every time, it follows that
\begin{equation} \label{eq:sensitivity_lemma}
\frac{\partial R}{\partial W_{ij}} 
= \int_0^T dt \frac{\delta R}{\delta W_{ij}(t)} 
= \int_{0}^T dt \frac{\delta R}{\delta g_i(t)}s_{ij}(t)
\end{equation}
We call this the {\em sensitivity lemma}, because it relates the sensitivity of $R$ to changes in $W_{ij}$ with the sensitivity to changes in $g_i(t)$.
The implication of the lemma is that
{\em dynamic} perturbations of the variables $g_i(t)$ can be used to instruct modifications of the {\em static} parameters $W_{ij}$. 

\noindent{\bf Gradient estimation}\
In order to estimate $\delta R/\delta g_i(t)$ 
suppose that Eq.\ (\ref{eq:Linear}) is perturbed by a fluctuating white noise,
\begin{equation}\label{eq:Perturbed}
g_i(t) = \sum_j W_{ij}s_{ij}(t)+\xi_i(t)
\end{equation}
The white noise satisfies $\langle\xi_i(t)\rangle=0$ and $\langle\xi_i(t_1)\xi_j(t_2)\rangle=\sigma^2\delta_{ij}\delta(t_1-t_2)$, where the angle brackets denote a trial average.
For now, let's regard this perturbation as a mathematical device; its biological interpretation will be discussed later. 

To show that $\delta R/\delta g_i(t)$ can be estimated from the covariance of 
$R$ and the perturbation $\xi_i(t)$, 
use the linear approximation $R-R_0 \approx \int_0^T dt \sum_k({\delta R}/{\delta g_k(t)}) \xi_k(t),$
which is accurate when the perturbations $\xi_i(t)$ are small. Here
$R_0$ is defined as $R$ in the absence of any perturbations, $\xi=0$. 
Since the perturbations are white noise, it follows that
\begin{equation}\label{eq:Correl}
\langle (R-R_0)\xi_i(t) \rangle \approx \sigma^2\frac{\delta R}{\delta g_i(t)}  
\end{equation}

Because $\langle \xi \rangle = 0$, the baseline $R_0$ may be replaced by any quantity that is uncorrelated with the perturbations of the current trial. 
For example, choosing $R_0=0$ leaves Eq.\ (\ref{eq:Correl}) valid.  However, baseline subtraction can have a large effect on the variance of the estimate (\ref{eq:Correl}) when based on a finite number of trials \cite{Dayan90}. Thus a good choice of baseline can decrease learning time, sometimes dramatically.


If the covariance relation of Eq.\ (\ref{eq:Correl}) is combined with the sensitivity lemma Eq.\ (\ref{eq:sensitivity_lemma}), it follows that
\begin{equation}\label{eq:correl}
\sigma^2\frac{\partial R}{\partial W_{ij}} \approx \int_0^T dt \langle (R-R_0)\xi_i(t) \rangle s_{ij}(t).  
\end{equation}

\noindent{\bf Synaptic learning rule} 
Equation (\ref{eq:correl}) suggests the following stochastic gradient learning procedure.
At each synapse the purely local {\em eligibility trace} 
\begin{equation}\label{eq:Eligibility}
e_{ij} = \int_0^T dt\, \xi_i(t)s_{ij}(t)
\end{equation}
is accumulated over the trajectory. 
At the end of the trajectory, the synaptic weight is updated according to: 
\begin{equation}\label{eq:Modification}
\Delta W_{ij} = \eta (R-R_0) e_{ij} 
\end{equation}
The update $\Delta W_{ij}$ fluctuates because of the randomness in the perturbations. On average, the update points in the direction of the gradient, because it satisfies 
$\langle\Delta W_{ij}\rangle \propto \partial R/\partial W_{ij}$, according to
Eq.\ (\ref{eq:correl}). This means that the learning rule of Eq.\ (\ref{eq:Modification}) is stochastic gradient following.

We note one subtlety in the derivation: In Eq.\ (\ref{eq:Eligibility}) the synaptic variables $s_{ij}(t)$ are defined in the presence of perturbations, while in the sensitivity lemma, they are defined for $\xi=0$.  In the linear approximations above, this discrepancy leads to a higher-order correction that is negligible for small perturbations.

\noindent{\bf Biological interpretation}
According to the above, synaptic weight gradients of $R$ can be estimated using conductance perturbations $\xi_i(t)$. Could this mathematical trick be used by the brain? In the actor-critic terminology of reinforcement learning \cite{Suttonbook98}, one can imagine that the neurons of one brain area (the ``actor'') drive actions that are assessed by another brain area (the ``critic''), which in response issues a global, scalar reinforcement signal $R$ to the actor (Fig. 1). A novel feature of our rule is that in addition to its regular synapses $W_{ij}$, the actor would receive a special class of ``empiric'' synapses from another hypothesized part of the brain (the ``experimenter''), which perturb the actor from trial to trial. Each plastic synapse locally computes and stores its scalar eligibility and multiplies this with $R$ to undergo modification. This idea is developed in detail elsewhere in a model of birdsong learning \cite{Fiete05bird,Fiete03}, resulting in concrete, nontrivial predictions for synaptic plasticity in the brain. 

Note that if the perturbation $\xi_i(t)$ is a synaptic conductance, its mean value $\langle\xi_i\rangle$ must be positive. Then the linear approximations above are expansions about the mean conductance $\xi_i(t)=\langle\xi_i\rangle$, rather than $\xi_i(t)=0$. As a result, $\xi_i(t)$ must be replaced by the zero-mean fluctuation $\delta\xi_i(t)=\xi_i(t)-\langle\xi_i\rangle$ in the eligibility trace. In addition, the fluctuations $\delta\xi_i(t)$ will not be truly white, but will have a correlation time set by the time constant of the synaptic currents. However, if this correlation time is short relative to the time scale of variation in ${\delta R}/{\delta g_i(t)}$, then the gradient estimate Eq.\ (\ref{eq:Correl}) should still be accurate.

Accurate gradient estimation requires that the eligibility trace filter out the mean conductance $\langle\xi_i\rangle$ of the empiric synapse. This  operation is  biologically plausible, and can be implemented by a simple time average at every ``actor'' neuron, if the empiric synapses are driven at a constant or very slowly varying rate.

By contrast, other proposals for stochastic gradient learning typically require individual neurons to keep track of and filter out a {\it time-varying} average vector of neural or synaptic activity within each trial, which seems rather complex.  The added complexity arises because these proposals are based on fluctuations in network dynamics caused by stochasticity intrinsic to neurons \cite{Barto85,Williams92,Xie04} or synapses \cite{Seung03} in the actor network; thus, the average perturbation is a function of the network trajectory and is time-varying. Our algorithm avoids this complexity, because the fluctuations are injected by an extrinsic source, and are therefore independent 
of the network trajectory. Our approach has the additional advantage that the degree of exploration in the actor can be modified independently of activity in the actor. 

\noindent{\bf Generalization to excitatory and inhibitory synapses}\
Above we assumed that all synapses have the same reversal potential. But neurons may receive both excitatory and inhibitory synapses, which have different reversal potentials. The unmodified learning rule allows both synapse types to perform gradient following if there are two types of empiric synapses per neuron: an excitatory empiric synapse used to train the excitatory synapses, and an inhibitory empiric synapse used to train the inhibitory synapses. But if there is only one empiric synapse per neuron, then for both types of synapses to perform gradient following, the rule must be modified.  Let $E_{ij}$ and $E_{\xi,i}$ be the reversal potentials of the regular $i\leftarrow j$ synapse and of the empiric synapse onto the $i$th actor neuron, respectively. Then we obtain a generalized sensitivity lemma:
\begin{equation} \label{eq:gen_sensitivity_lemma}
\frac{\partial R}{\partial W_{ij}} 
= \int dt  \ a_{ij}(t) \frac{\delta R}{\delta g_i(t)}  s_{ij}(t)
\end{equation}
where
\begin{equation}\label{eq:Ratio}
a_{ij}(t)=\frac{V_i(t)-E_{ij}}{V_i(t)-E_{\xi,i}}
\end{equation}
is the ratio of the synaptic driving force at the $i\leftarrow j$ synapse to
the driving force of the empiric synapse at neuron $i$. The stochastic
gradient learning rule remains $\Delta W_{ij}=\eta(R-R_0) e_{ij}$, but with 
modified eligibility trace
\[
e_{ij}= \int_{0}^T a_{ij}(t) \xi_i(t)s_{ij}(t),
\]
For synapses with the same reversal potential as the empiric synapse, $a_{ij}(t)=1$, returning the original learning rule. Even for synapses of the opposite variety, the sign of $a_{ij}$ does not change with time because neural voltage is constrained to stay between the inhibitory and excitatory reversal potentials $V_I$ and $V_E$ ($V_I \leq V_i(t)\leq V_E$), and  $E_{\xi,i}, E_{ij} \in \{ V_I, V_E\}$. Nevertheless, for these synapses of the opposite variety, the term $a_{ij}(t)$ adds complexity to the simple learning rule and reduces its biological plausibility. 

\noindent{\bf Generalization to multicompartmental model neurons} \ Suppose the
model neuron is not isopotential, but has several dendritic compartments. Then it can be trained without modification of the learning rule by using a separate empiric synapse for each compartment. Alternatively, a single empiric synapse could be used for the whole neuron, but with the introduction of 
complexities in the learning rule similar to the $a_{ij}(t)$ factor of Eq.\ (\ref{eq:Ratio}).

\noindent{\bf Technical issues}\  Our synaptic learning rule performs stochastic gradient following, and therefore shares the virtues and deficiencies of all methods in this class \cite{Pearlmutter95}. For example, it is possible to become stuck at a local optimum of the objective function. The stochasticity of the gradient estimation may allow some small probability of escape, but there is no guarantee of finding a global optimum.

The derivation of our learning rule in particular, and of gradient rules in general, depends on the smoothness assumption that $R$ is a differentiable function of the synaptic weights.  
But $R$ depends on $W_{ij}$ through the spiking activity of the actor network, and spiking neurons typically exhibit threshold spike- or no-spike behaviors, so one might worry that $R$ is discontinuous.  However, because either the amplitude or the latency of neural spiking varies continuously as a function of input near threshold \cite{Rinzel89}, there is typically no true discontinuity. 


\noindent{\bf Comparison with previous work} If the perturbation $\xi_i(t)$ is {\em Gaussian} white noise, then our synaptic learning rule can be included as a member of the REINFORCE class of algorithms \cite{Williams92}. With Gaussian white noise we can use the REINFORCE formalism to prove that our learning rule performs stochastic gradient ascent on $R$ without assuming that the perturbations are small, because linear approximations are not used. In contrast, our present derivation does not require the perturbations to be Gaussian, but assumes they are approximately white, and of small amplitude. The REINFORCE theory too could be used for non-Gaussian $\xi_i(t)$, if $\xi_i(t)$ is drawn i.i.d.\ from a smooth probability density function (PDF). However, the resulting learning rule will be different than ours. Further, the assumption of smoothness of the PDF can seriously limit the applicability of the REINFORCE theory: for example, a $\xi$ generated by filtering a random spike train cannot be treated by REINFORCE. 

The sensitivity lemma allows us to derive rules for synaptic gradient learning based on perturbations of other quantities not directly related to the synaptic parameters. Versions of the sensitivity lemma have appeared in the literature for nonspiking feedforward networks, and been used to estimate the gradient by serially perturbing one neuron at a time (node perturbation) \cite{Andes90,LeCun89}. Our version of the sensitivity lemma is more general, because it is applicable to learning trajectories in recurrent networks, via parallel perturbation of multiple neurons. Most importantly, we have shown how to use it to derive a biologically plausible rule for gradient learning in spiking networks. 
%

\noindent{\bf Acknowledgments}  For comments on the manuscript, the authors are grateful to Y. Loewenstein and U. Rokni.  I.F. acknowledges funding from NSF PHY 99-07949.  

\noindent{\bf APPENDIX: Stochastic networks}\ Above the network dynamics and reinforcement $R$ were assumed to be deterministic.  Both elements can be made stochastic, as outlined below. Consider the case of discrete time (continuous time is a limiting case). 
The network generates a trajectory $\Omega = \{\Omega(0),\Omega(1),\ldots,\Omega(T)\}$ from a probability density $P_W(\Omega)$. Suppose each trajectory is generated by drawing an initial condition $\Omega(0)$ from some probability density and then drawing $\Omega(1)$ through $\Omega(T)$ from a Markov process with transition probability $P_W(\Omega(t)|\Omega(t-1))$.  The assumption of Markov transition probabilities is compatible with most spiking neural network models.  The network receives
reinforcement $R$ from the conditional density $P(R|\Omega)$. Since the network is parametrized by $W$, the expected reward
\begin{equation}\label{eq:ExpectedReward}
\langle R\rangle = 
\int R P(R|\Omega) P_W(\Omega)dR\cd \Omega
\end{equation}
is a function of $W$. We assume that the transition probability depends on the weights $W$ through 
\begin{equation}\label{eq:TransitionConductance}
P_W(\Omega(t)|\Omega(t-1))=f(g_1(t),\ldots,g_N(t))
\end{equation}
where as before
\begin{equation}\label{eq:Linear2}
g_i(t) = \sum_j W_{ij}s_{ij}(t-1)
\end{equation}
The transition probability depends on all the dynamical variables in $\Omega(t)$, although they have been suppressed for notational simplicity in Eq.\ (\ref{eq:TransitionConductance}).  As before, the important mathematical property here is the linearity of Eq.\ (\ref{eq:Linear2}), which is embedded inside a nonlinear system. The sensitivity lemma takes the form:
\begin{equation}
\frac{\partial \langle R \rangle}{\partial W_{ij}} 
= \sum_{t=1}^T \frac{\partial}{\partial g_i(t)}\langle R s_{ij}(t-1) \rangle
\end{equation}  
The sensitivity lemma shows that the appropriate change in the weight of a synapse is not given by the covariance of its activity with reinforcement (as might be naively expected), but is instead given by the derivative with respect to $g_i(t)$ of this covariance. As before, the proof of the sensitivity lemma involves comparing derivatives of the transition probabilities taken with respect to $W_{ij}$ and $g_i(t)$, without actually performing either differentiation. Note that REINFORCE requires the stronger condition that the log probability be differentiable. For small perturbations $\xi_i(t)$, this sensitivity lemma leads us again to the gradient learning rule of Eqns. (\ref{eq:Eligibility}-\ref{eq:Modification}), now valid for fully stochastic networks. 

\bibliographystyle{revtex}

\begin{thebibliography}{10}
\providecommand*{\bibinfo}[2]{#2}
\providecommand*{\eprint}[1]{#1}
\providecommand*{\url}[1]{#1}

\bibitem{LeCun98a}
\bibinfo{author}{Y.~LeCun {\emph{et al.}}}, 
  \bibinfo{journal}{Proc IEEE} \bibinfo{volume}{\textbf{86(11)}},
  \bibinfo{pages}{2278} (\bibinfo{date}{1998}).

\bibitem{Widrow90}
\bibinfo{author}{B.~Widrow} and \bibinfo{author}{M.~Lehr},
  \bibinfo{journal}{Proc IEEE} \bibinfo{volume}{\textbf{78}}(9),
  \bibinfo{pages}{1415} (\bibinfo{date}{1990}).

\bibitem{Rumelhart86}
\bibinfo{author}{D.~Rumelhart {\emph{et al.}}}, 
in \bibinfo{editors}{D.~Rumelhart and J.~McClelland}, eds., \emph{Parallel Distributed Processing}
  (\bibinfo{publisher}{MIT Press}, \bibinfo{year}{1986}).

\bibitem{Markram96}
\bibinfo{author}{H.~Markram} and \bibinfo{author}{M.~Tsodyks},
  \bibinfo{journal}{Nature} \bibinfo{volume}{\textbf{382}}(6594),
  \bibinfo{pages}{807} (\bibinfo{date}{1996}).

\bibitem{Thomson94}
\bibinfo{author}{A.~Thomson} and \bibinfo{author}{J.~Deuchars},
  \bibinfo{journal}{Trends Neurosci.} \bibinfo{volume}{\textbf{17}},
  \bibinfo{pages}{119} (\bibinfo{date}{1994}).

\bibitem{Dembo90}
\bibinfo{author}{A.~Dembo} and \bibinfo{author}{T.~Kailath},
  \bibinfo{journal}{IEEE Trans on Neural Networks}
  \bibinfo{volume}{\textbf{1}}(1), \bibinfo{pages}{58} (\bibinfo{date}{1990}).

\bibitem{Seung03}
\bibinfo{author}{H.~Seung}, \bibinfo{journal}{Neuron.}
  \bibinfo{volume}{\textbf{40}}(6), \bibinfo{pages}{1063} (\bibinfo{date}{2003}).

\bibitem{LeCun89}
\bibinfo{author}{Y.~LeCun {\emph{et al.}}}, 
in \bibinfo{editors}{D.~Touretzky}, ed.,
  \emph{Adv Neural Info Proc Sys 1}, \bibinfo{pages}{141} 
  (\bibinfo{date}{1989}).

\bibitem{Werfel05}
\bibinfo{author}{J.~Werfel}, \bibinfo{author}{X.~Xie}, and
  \bibinfo{author}{H.~Seung}, \bibinfo{journal}{Neural Comp}
  \bibinfo{volume}{\textbf{17(12)}}, \bibinfo{pages}{2699}
  (\bibinfo{date}{2005}).

\bibitem{Dayan90}
\bibinfo{author}{P.~Dayan}, in \bibinfo{editors}{D.~Touretzky {\emph{et al.}},
}, eds., \emph{Proc Connectionist Models Summer
  School} (\bibinfo{publisher}{Morgan Kaufmann}, \bibinfo{year}{1990}).

\bibitem{Suttonbook98}
\bibinfo{author}{R.~Sutton} and \bibinfo{author}{A.~Barto},
  \bibinfo{title}{\emph{Reinforcement learning: An introduction}}
  (\bibinfo{publisher}{MIT Press}, \bibinfo{year}{1998}).

\bibitem{Fiete03}
\bibinfo{author}{I.~Fiete}, 
Ph.D. thesis, Harvard University
  (\bibinfo{date}{2003}).

\bibitem{Fiete05bird}
\bibinfo{author}{I.~Fiete} and \bibinfo{author}{H.~Seung},
  \bibinfo{journal}{Submitted}  (\bibinfo{date}{2005}).

\bibitem{Barto85}
\bibinfo{author}{A.~G. Barto} and \bibinfo{author}{P.~Anandan},
  \bibinfo{journal}{IEEE Trans on Systems, Man, and Cybernetics}
  \bibinfo{volume}{\textbf{15}}(3), \bibinfo{pages}{360}
  (\bibinfo{date}{1985}).

\bibitem{Williams92}
\bibinfo{author}{R.~Williams}, \bibinfo{journal}{Machine Learning}
  \bibinfo{volume}{\textbf{8}}, \bibinfo{pages}{229} (\bibinfo{date}{1992}).

\bibitem{Xie04}
\bibinfo{author}{X.~Xie} and \bibinfo{author}{H.~Seung}, \bibinfo{journal}{Phys
  Rev E} \bibinfo{volume}{\textbf{69}},
  \bibinfo{pages}{041909} (\bibinfo{date}{2004}).
  
\bibitem{Pearlmutter95}
\bibinfo{author}{B.~Pearlmutter}, \bibinfo{journal}{IEEE Trans on
  Neural Networks} \bibinfo{volume}{\textbf{6}}(5), \bibinfo{pages}{1212}
  (\bibinfo{date}{1995}).

\bibitem{Rinzel89}
\bibinfo{author}{J.~Rinzel} and \bibinfo{author}{B.~Ermentrout}, in
  \bibinfo{editors}{C.~Koch and I.~Segev}, eds., \emph{Methods in Neuronal
  Modelling: From synapses to Networks} (\bibinfo{publisher}{MIT Press},
  \bibinfo{year}{1989}).

\bibitem{Andes90}
\bibinfo{author}{D.~Andes {\emph{et al.}}}, 
in \emph{IJCNN-90-WASHDC: International Joint Conference on Neural Networks},
 (\bibinfo{publisher}{Lawrence Erlbaum Associates}, \bibinfo{year}{1990}).

\end{thebibliography}

\end{document}